\documentclass[12pt]{article}
\usepackage{epsfig}
\begin{document}
\title{Crossover from Reptation to Rouse dynamics in the Extended Rubinstein-Duke Model}
\author{Andrzej Drzewi\'nski \\
Institute of Physics, University of Zielona G\'ora, \\
Prof. Z. Szafrana 4a, 65-516, Zielona G\'ora, Poland\\*[4mm]
J.M.J. van Leeuwen \\
Instituut-Lorentz, University of Leiden, P.O.Box 9506, \\
2300 RA Leiden, the Netherlands} 
\maketitle

\begin{abstract}
The competition between reptation and Rouse Dynamics is incorporated in 
the Rubinstein-Duke model for polymer motion by extending it with sideways motions, 
which cross barriers and create or annihilate hernias.
Using the Density-Matrix Renormalization-Group Method as solver of the Master Equation, the renewal time and the 
diffusion coefficient are calculated as function of the length of the chain and the strength 
of the sideways motion.
These new types of moves have a strong and delicate influence on the asymptotic 
behavior of long polymers. The effects are
analyzed as function of the chain length in terms of effective exponents and crossover 
scaling functions.
\end{abstract} 

PACS {61.25.Hk, 05.10.-a, 83.10.Kn}

\section{Introduction}

A dilute solution of linear polymers in a gel provides the ideal case for reptation. The
gel is a rigid network of obstacles which forces the polymer to find its way slithering
through the maze. Effectively the polymer moves inside a tube of pores which changes only
by growing and shrinking at the ends of the tube. It is not a great step to 
replace the network of the gel by a regular lattice. The regularity of the lattice still 
keeps the motion of the polymer random, because the ends randomly leave or enter 
cells of the lattice. A big step is the
reduction of the motion to a stochastic process of hopping units. It certainly can not
be justified on the level of monomers, because neighboring monomers are strongly correlated
in their motion. For this purpose the notion of reptons has been introduced: blobs of 
monomers of the size of the correlation length \cite{deGennes}. Seeing the polymer as a sequence of 
reptons, permits to consider the units of motion as uncorrelated, with the only  
proviso that they do not separate too far in order to preserve the integrity of the polymer.

Rubinstein \cite{Rubinstein} designed an elementary model for reptation, 
as a chain of slack and taut links connecting the reptons. A slack link describes
two successive reptons in the same cell and a taut link two in nearest neighbor cells. 
By allowing only these configurations of reptons and only moves between them, 
a simple model for reptation results. Duke \cite{Duke} enriched the model 
by biasing the hopping of reptons by an external field, thus modeling the experimental 
situation of gel electrophoresis. Of course the model misses important aspects of polymer 
dynamics, such as the hydrodynamic interactions and even more importantly the requirement 
of self-avoidance, which influences the universal properties \cite{Doi,Viovy}. 
One can incorporate self-avoidance in the model, which makes however the analysis an order
of magnitude more difficult. An excuse for leaving out this aspect is that mutual exclusion
is for reptons less severe than for monomers, because the reptons are loosely packed
blobs of monomers which can interpenetrate each other.  

The eternal dilemma is to choose between being realistic and keeping it simple. The 
chemical culture opts for being realistic and deals with specific properties, the physical
culture leans towards simplicity and aims at generic properties. It was de Gennes' 
contribution to polymer physics to show that properties of long polymers do not depend
on the specific composition, in particular the dependence on the length of the polymers
is governed by universal exponents. In this sense the Rubinstein-Duke (RD) model has 
shown to catch the essential physics of reptation in spite of the crude approximations made.

We will approach the problem from the physical perspective
and deal with the universal properties of the RD model, but investigate a richer 
class of motions than treated sofar.
In the standard RD model only interchange of slack and taut links is permitted. It means
that the length stored in a slack link moves in the direction of a taut link, interchanging 
the slack and taut link. A move that would perfectly fit in the spirit of the RD model, 
is the change of two consecutive slack links into two taut links.
It corresponds to three reptons in the same cell of which the middle one escapes to
a neighboring cell. Such a move does not cross a barrier.
There was a practical reason to exclude this possibility, because it
destroys the dimensional reduction, which was pointed out by Duke. From
a physical point of view, the formation of ``hernias'' does not seem to
influence the universal properties and in this paper we investigate this issue.
Another optional move is the interchange of two taut links. As we will see this means the 
crossing of a barrier by the chain. If the barriers, posed by the obstacles were 
infinitely high, these processes would be strictly forbidden. But barriers are not
perfect and therefore it is worthwile to investigate the influence of finite barriers.
Moreover we will see that barrier crossing in the RD model 
does not lead to much change, but the combination with hernia creation and 
annihilation has the drastic effect of crossing over from reptation to Rouse dynamics.

Since the standard RD model already does not permit an exact solution, we have to rely
on numerical methods to analyze the extended model. The most common method is simulation
of the system but this is less suited for our goal, because the crossover between the
two types of dynamics occurs for rather long chains which are hard to simulate
accurately. We will employ the technique of finite size analysis, which requires
very accurate data to be successful.
In this paper we use a method, based on the analogy between the Master 
Equation and the Schrodinger equation, by which the temporal evolution of the probability 
distribution of the chain configurations corresponds to the evolution of the wave function. 
The Master operator corresponds to the Hamiltonian of a one-dimensional spin chain, for which
the very efficient Density-Matrix Renormalization-Group Method (DMRG) has been designed by White \cite{White}.
The model remains a one-dimensional quantum problem, irrespective the lattice in which
it is embedded, because the chain itself is a linear structure. Application of the 
DMRG method to the chain dynamics on the lattice is by now standard, but to perform 
successfully calculations  in a $3d$ embedding lattice requires an optimal use of
the symmetries of the model in order to keep the basis set of states to a practical 
size. In the appendix we outline how we exploit the symmetries of the $3d$ lattice.

In the next section we describe the extension of the RD model and the 
corresponding Master Equation. 
We focus on two properties: the renewal time $\tau$ and the diffusion coefficient 
$D$ and determine them directly from the Master Operator. 
The renewal time is the time needed for the chain to assume a new configuration,
which has no memory of the original one. It is found from the gap in the spectrum of the
Master Operator. The Master Equation always has a trivial eigenvalue 0, corresponding
to the stationary state. Any other initial state ultimately decays towards the stationary
state and the slowest relaxation time (the inverse of the gap) is the renewal time. 
The gap decays with a negative power $z$ of the length $N$ of the chain, such that 
$\tau \sim N^z$. The zero field diffusion coefficient $D$ is related to the drift velocity 
in a weak driving field and decays as a power $N^{-x}$. The approach to this 
asymptotic behavior is the main issue of this paper.

Due to the new types of hopping, the dimensionality $d$ of the embedding lattice plays a 
non-trivial role. We report, for the first time, on calculations in $d=3$. They became
possible through subtile use of the symmetries of the model, which are discussed in the 
appendix. The subsequent sections contain the results for the renewal exponent $z$ and 
the crossover functions, which describe the data for all lengths $N$ and
strengths of the transition rates for barrier crossing and hernia creation/annihilation.
The results for the diffusion coefficient $D$ and its exponent $x$ are calculated from a
linearization of the Master Equation with respect to the driving field. The exponents $z$ 
and $x$ are linked through the mean square displacement of the wandering chains.
In the discussion we comment on the results and explain why the crossover in gels is
different from that in polymer melts.

Paessens and Sch\"utz \cite{paessens} have also extended the RD model by including "constraint
release" in the hopping rates. In our language this is a mix of hernia creation/annihilation and
barrier crossings. We comment their calculation in the discussion.

Earlier \cite{DvL1} we have performed a similar investigation for the cage model (in $d=2$), with
similar conclusions as the present study. Investigation of the RD model elucidates in how far
the crossover is model independent. 
 
\section{The model}

The model consists of a chain of $N+1$ reptons located in the cells of a (hyper)cubic
lattice. They are connected by $N$ links, labeled by ${\bf Y} = (y_1, \cdots, y_N)$.
The links take on the value $y_i =0$ (slack) or any of the $2d$ vectors which connect
a cell to its neighbors (taut). The corners of the squares in $d=2$ or the edges of the 
cubes in $d=3$ are barriers for the chain. The reptons can move in three ways: 
\begin{enumerate}
\item The standard RD moves. A repton between a slack and taut link moves to the 
neighboring cell thereby interchanging the slack and taut link. For these moves
no barriers have to be overcome. A move is illustrated in Fig. \ref{rdmove}. 
\begin{figure}[h]
\begin{center}
    \epsfxsize=10cm%\linewidth
    \epsffile{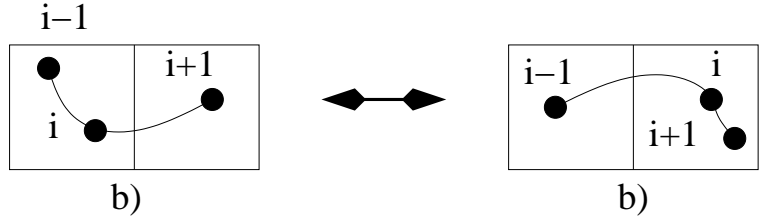}
    \caption{Rubinstein-Duke moves. Their rate is set to 1.}  
\label{rdmove}
\end{center}
\end{figure}
The strength of the hopping rate for RD moves sets the time scale and is therefore 
put equal to 1.
\item The barrier crossings, of which an example is shown in Fig. \ref{barrier}.
\begin{figure}[h]
\begin{center}
    \epsfxsize=10cm%\linewidth
    \epsffile{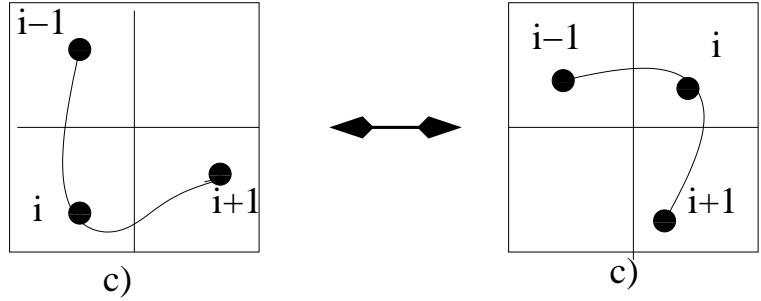}
    \caption{Barrier crossings with hopping rate $c$}  
\label{barrier}
\end{center}
\end{figure}
This is an interchange of two taut links connected to the same repton that jumps over 
the barrier. The strength of the transition rate for such a move is taken to be $c$.
\item The creation of a hernia is the change from two consecutive slack links of 
which the middle repton jumps to a neighboring cell. The annihilation is the reverse
process. An example is shown in Fig. \ref{hernia}.
\begin{figure}[h]
\begin{center}
    \epsfxsize=10cm%\linewidth
    \epsffile{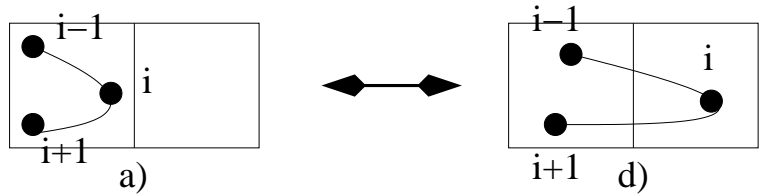}
    \caption{Hernia creation a) to d) and annihilation d) to a). Both processes go with
rate are $h$.}  
\label{hernia}
\end{center}
\end{figure}
Hernia creation and annihilation go with the rate $h$.
\end{enumerate}

The statistics of the model is governed by the Master Equation for the probability 
distribution $P ({\bf Y},t)$, where ${\bf Y}$ stands for the complete 
configuration $( y_1, \cdots, y_N)$. It has the general form
\begin{equation} \label{a1}
{\partial P({\bf Y},t) \over \partial t} =  \sum_{\bf Y'}\left[ W ({\bf Y} | {\bf Y}')
P ({\bf Y}',t) - W ({\bf Y}' | {\bf Y}) P({\bf Y},t)\right]
\equiv \sum_{\bf Y'} M({\bf Y},{\bf Y}') P({\bf Y}',t).
\end{equation}
The $W$'s are the transition rates of the possible motions that we have indicated in
the above list. 
The matrix $M$ combines the gain terms (in the off-diagonal elements) and the loss
terms (on the diagonal). $M$ is the sum of matrices, for each repton one
\begin{equation} \label{a2}
M({\bf Y},{\bf Y}') = \sum^N_{i=0} M_i({\bf Y},{\bf Y}'),
\end{equation}
where the sum runs over the reptons starting with the tail repton $i=0$ to the head
repton $i=N$. The internal reptons induce transitions between two configurations 
which differ in two consecutive $y_i$, the external (head and tail) repton change 
only $y_N$ viz. $y_1$. If we view the links as ``bodies'' the problem is equivalent
with a one-dimensional many body system with two body interactions between nearest
neighbors.  

The matrix $M$ is asymmetric because the transition rates are biased by a factor
$B=\exp(\epsilon /2)$ for the moves in the direction of the field and by $B^{-1}$ for
the reverse process. $\epsilon$ is a dimensionless parameter representing the 
strength of the field. $M$ is a stochastic matrix since the sum over each column vanishes. 
Therefore $M$ has an eigenvalue 0 and the corresponding right eigenvector is the 
probability density of the stationary state. All other eigenvalues are negative. 
The smallest in magnitude is the gap, giving the slowest decay to the stationary 
state and the inverse of the gap we take as the definition of the renewal time. 
The diffusion coefficient $D$ is calculated from an infinitesimal
small driving field. The field induces a drift $v_d$ and via the Einstein relation
\begin{equation} \label{a3}
D = {1 \over N} \,\left( {\partial v_d \over \partial \epsilon }\right)_{\epsilon=0}
\end{equation} 
the diffusion coefficient results. It is determined by expansion of the   
Master Equation in powers of $\epsilon$. 
\begin{equation} \label{a4}
{\cal M} = {\cal M}_0 + \epsilon {\cal M}_1 + \cdots , \quad \quad \quad 
P ({\bf Y}) = P_0 ({\bf Y}) + \epsilon P_1 ({\bf Y}) + \cdots 
\end{equation} 
which leads to the equations
\begin{equation} \label{a5}
{\cal M}_0 P_0 = 0 , \quad \quad \quad 
{\cal M}_0 P_1 = - {\cal M}_1 P_0.
\end{equation} 
The first equation is trivially fulfilled by a constant $P_0 ({\bf Y})$, since the matrix 
$M_0$ is symmetric and the right eigenvector becomes equal to the trivial left eigenvector.
The right hand side of the second equation is a known function of the configuration. Thus it
yields the linear perturbation $P_1$.  The drift velocity is an average over
the distribution \cite{Widom}. The linear term in $\epsilon$ of the drift velocity involves
the terms $P_0$ and $P_1$. With these terms we can calculate the linear term in 
$v_d$ and find with (\ref{a3}) the diffusion coefficient $D$.

\section{The Exponent $z$ for the Renewal Time}

The easiest way to obtain the exponent of the relation $\tau = N^z$ is to make a 
log-log plot and determine the slope. In a previous publication \cite{Carlon} it was 
shown that this is rather misleading for the present problem. A much more sensitive check 
is to compute local exponents $z_N$ according to
\begin{equation} \label{b1}
z_N = {\ln \tau(N+1) - \ln \tau(N-1) \over \ln (N+1) - \ln (N-1)} 
\simeq {d \ln \tau \over d \ln N},
\end{equation}
which gives $z$ as a function of the chain length $N$. The function $z_N$ is the basic 
ingredient for further analysis. 
\vspace*{1.3cm}

\begin{figure}[ht]
\begin{center}
    \epsfxsize=12cm%\linewidth
    \epsffile{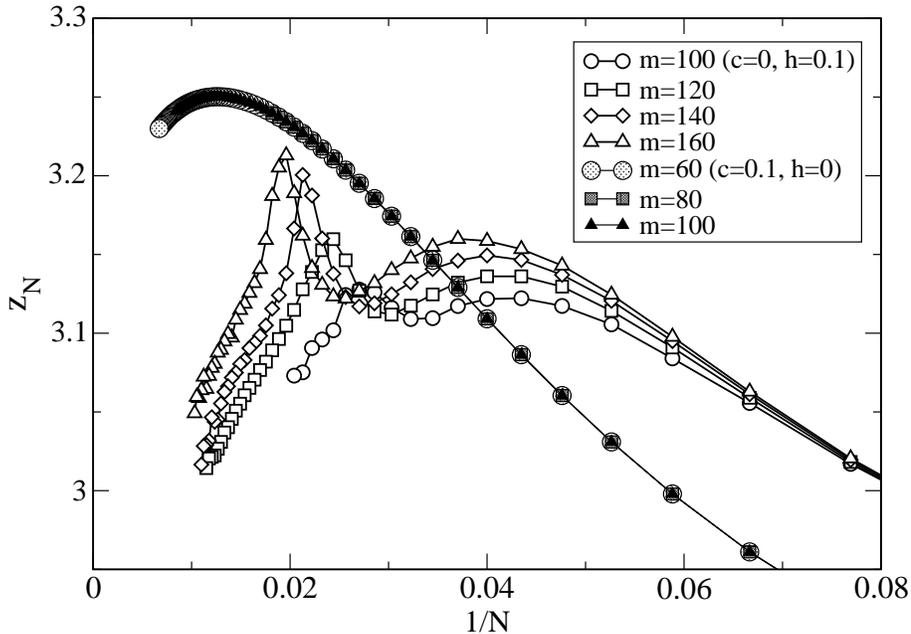}
    \vspace*{5mm}

    \caption{The influence of the basis size $m$ on the renewal time exponent for two combinations 
of parameters: $c=0, h=0.1$ and $c=0.1, h=0$. }  \label{convhernia}
\end{center}
\end{figure}
First we check how accurate it can be obtained from a DMRG calculation
of which the convergence is determined by the basis size $m$.  
In Fig. \ref{convhernia} we show examples of poor and excellent convergence. Poor
convergence occurs for the combination $c=0, h=0.1$, where we have no barrier crossings but 
substantial hernia creation/annihilation. For a length $N=20$ the size of the basis,
even as large as m=160, still has an influence and for longer chains this becomes stronger. 
Thus it is difficult to deduce from these data the effective exponent for chains longer than 
$N=20$. In the second combination $c=0.1, h=0$ the convergence is perfect for much smaller 
bases and for much longer chains, as the upper curve in the picture demonstrates. 
In fact the case with no barrier crossing at all is the only combination where convergence 
is a problem, as subsequent pictures will show. We blame the lack of convergence to the 
fact that hernia creation/annihilation
without some barrier crossing leads to a large weight for configurations with many hernias
and therefore to a short end-to-end distance of the chain. These are a-typical configurations
and the DMRG procedure has difficulty to find an adequate basis to represent the gap state. 

Some other noticeable points are:
\begin{enumerate}
\item The curves have still not reached the asymptotic value for the values of $N$ of the 
order of 100. Thus a log-log plot would suggest a higher value than the reptation value 3. 
This slow approach to the asymptotic value has been identified as the main 
reason for the discrepancy between the theoretical reptation exponent z=3 and the measured 
higher values \cite{Carlon}. We get a better grip on the asymptotics when we discuss
the crossover. 
\item The curves do not indicate a tendency towards the Rouse exponent $z=2$.
This illustrates the point made in the introduction that the two mechanisms have to assist 
each other, before deviations from reptation occur.
\end{enumerate}
\begin{figure}[ht]
\begin{center}
    \epsfxsize=12cm%\linewidth
    \epsffile{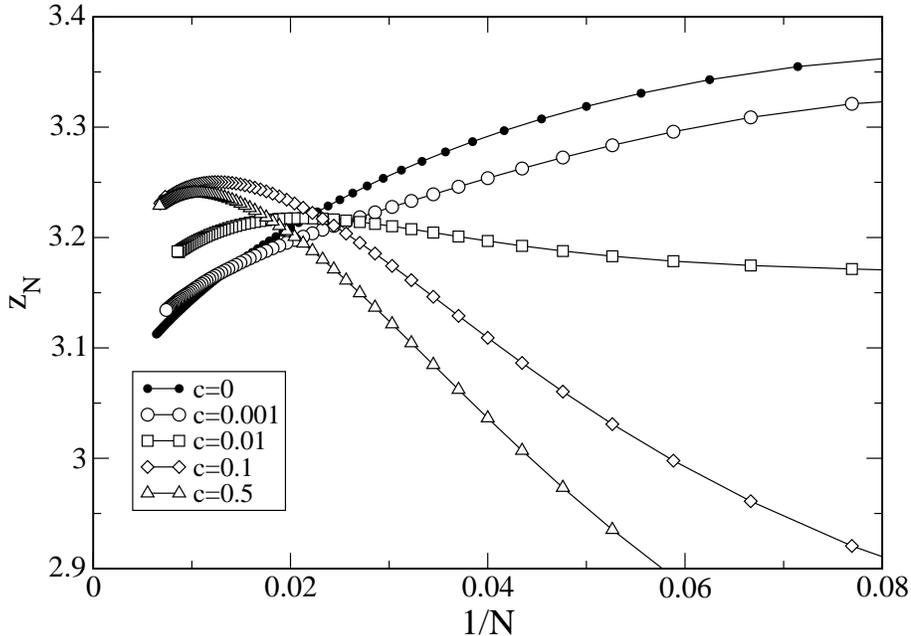}
       \vspace*{5mm}

    \caption{The renewal time exponent $z_N$ for $h=0$ and a set of values $c$. } \label{lackbar}
\end{center}
\end{figure}
The next Fig. \ref{lackbar} shows a set of curves for $h=0$ and a set of values $c$. Note 
that the effective exponent is quite sensitive to the value of $c$, but all curves do not 
show reptative behavior as was mentioned earlier for the $N$ dependence of a single
point in parameter space. For larger values of $c$, a maximum in the effective
exponent $z_N$ seems to develop for larger and larger $N$. 
As the maximum can easily be interpreted as a saturated asymptotic value in a log-log plot, 
these corrections to scaling, still present for very long chains, are very important
for assessing the correct asymptotic behavior. This feature makes it neccesary to do a
finite size analysis in order to get a grip on the region in $N$ where the behavior changes. 

The standard argument is that hernia creation/annihilation does not change the reptative 
character of the chain motion because they leave the backbone of the chain invariant,
which is the collection of taut links after the chain has been successively stripped from its hernias. 
The backbone only changes by refreshment at the ends of the chain. Also barrier crossing
seems to be, as a single mechanism, ineffective. It changes the backbone, but not the number of taut
links in a certain direction, since in a barrier crossing taut links are only interchanged 
in position along the chain. We may call the properties of the chain, which do not change 
by internal motion, ``quasi invariants''. 
So one needs both a non-zero $c$ and a non-zero $h$ to remove these quasi invariants. 
The cooperation of hernia creation/annihilation and barrier crossing is a intricate mechanism.
Therefore we concentrate first on the situation that one of them has a finite strength and the 
other becomes small. 

\section{Simple Crossover}

\begin{figure}[ht]
\begin{center}
    \epsfxsize=12cm%\linewidth
    \epsffile{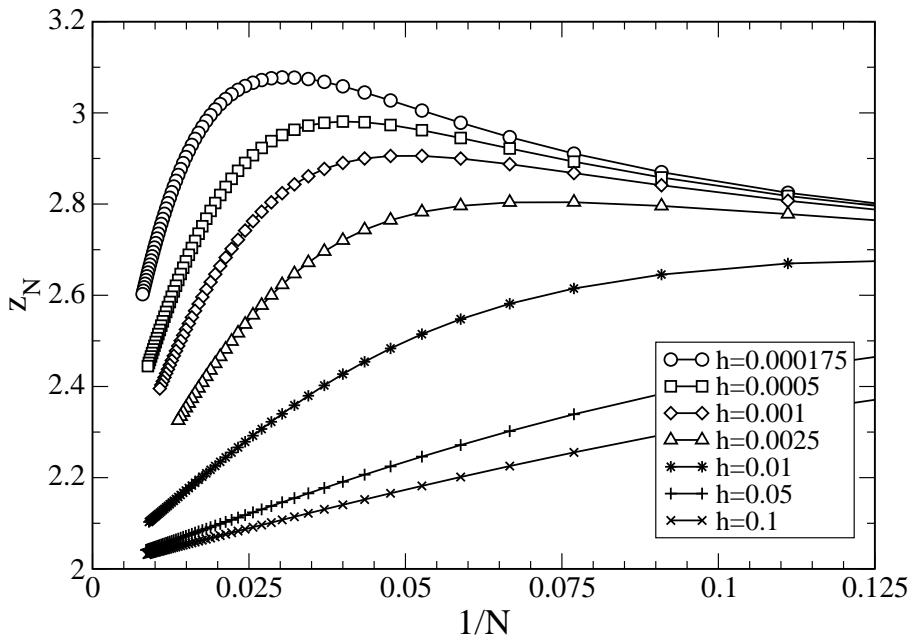}
    \vspace*{5mm}

    \caption{The renewal time exponent $z_N$ for $c=0.1$ and a set of values $h$. }\label{simcrossc1}
\end{center}
\end{figure}
\begin{figure}[ht]
\begin{center}
    \epsfxsize=12cm%\linewidth
    \epsffile{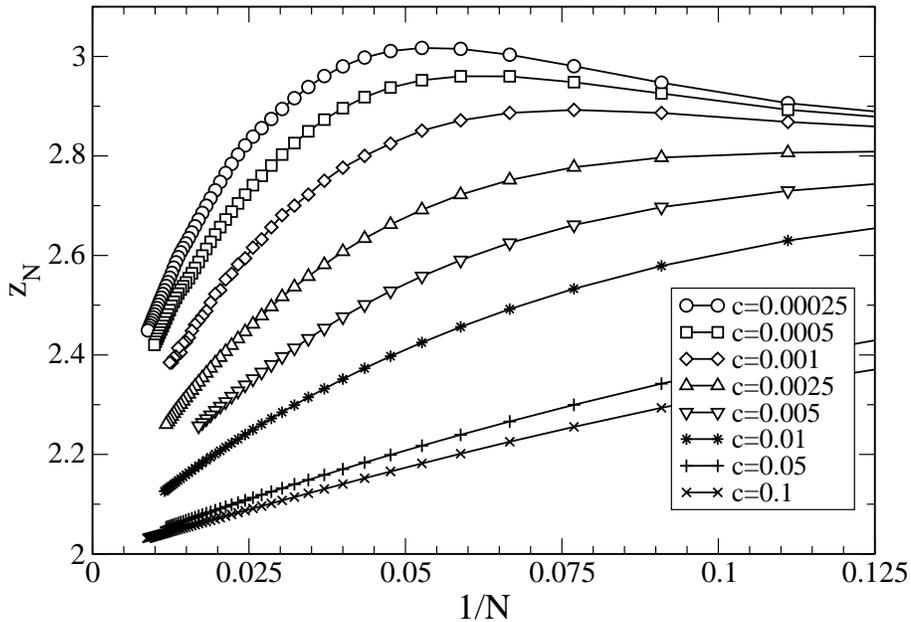}
   \vspace*{5mm}

    \caption{The renewal time exponent $z_N$ for $h=0.1$ and a set of values $c$. }\label{simcrossh1}
\end{center}
\end{figure}

The case where one of the two, $c$ or $h$, is fixed at a finite value, 
gives a simple crossover from reptation to Rouse dynamics when the chain grows with $N$. 
As example consider first a fixed value $c=0.1$ and $h$ varying and small, 
for which the local exponent $z_N$ is given in Fig. \ref{simcrossc1}. 
The reverse situation is plotted in Fig. \ref{simcrossh1} with $z_N$ for $h=0.1$ and a set of 
values of $c$.
The two figures are strikingly similar. For small values of the parameter $c, (h)$ the chain 
seems to show reptative behavior but turns over towards the Rouse exponent $z=2$ for 
longer chains. It is remarkable that even in Fig. \ref{simcrossh1} the values for very 
small $c$ show this trend, while we know from the previous section that for $c=0$ the 
calculation is poorly convergent.
\begin{figure}[ht]
\begin{center}
    \epsfxsize=12cm%\linewidth
    \epsffile{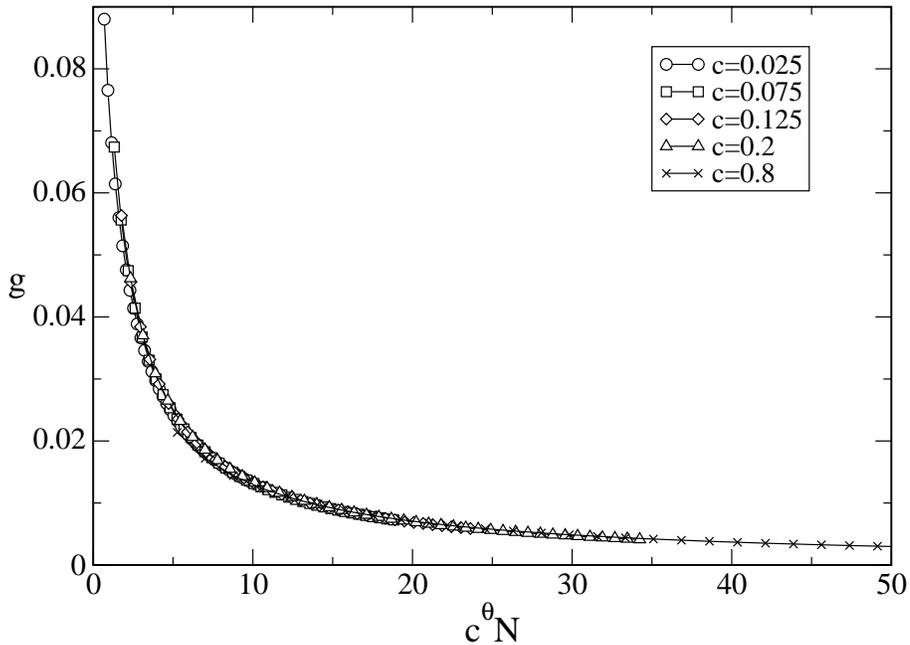}
   \vspace*{5mm}

    \caption{The crossover scaling function $g(x)$ for $h=0.5$ and $\theta=0.58$.}  \label{scalingc5}
\end{center}
\end{figure}

Anticipating the asymptotic values of the two regimes, 
the following representation is adequate for the renewal time (for fixed $h$).
\begin{equation} \label{e1}
\tau (N, c) = N^3 g (c^\theta N).
\end{equation} 
The idea is that all curves of e.g. Fig. \ref{simcrossh1} are represented by a single 
curve $g(x)$. Thus we have plotted in Fig. \ref{scalingc5} the data for $\tau N^{-3}$ as 
function of $c^\theta N$ for the fixed value $h=0.5$ and varying $c$, with an assumed value 
$\theta=0.58.$ 
\begin{figure}[ht]
\begin{center}
    \epsfxsize=12cm%\linewidth
    \epsffile{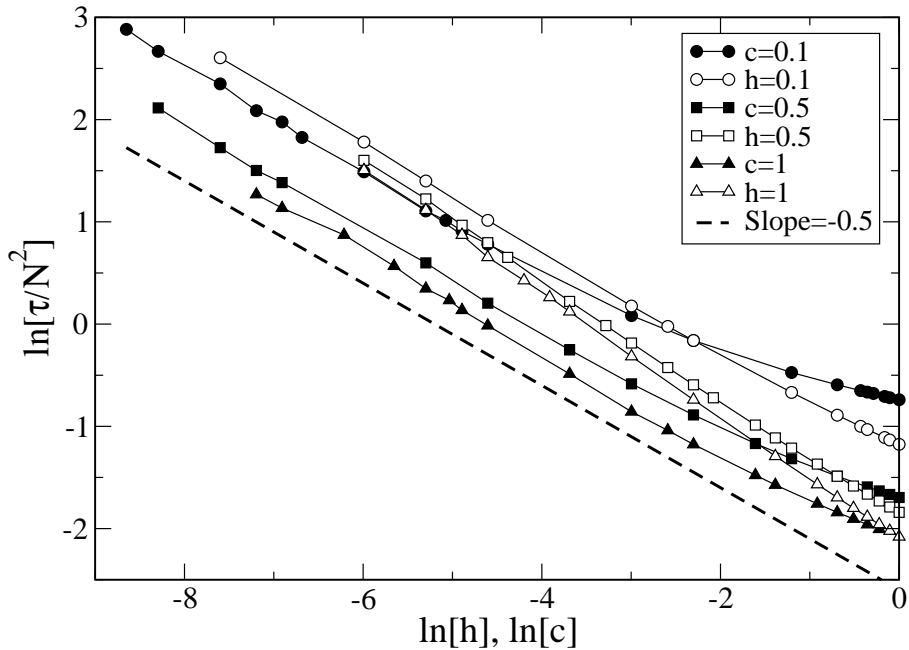}
   \vspace*{5mm}

    \caption{The crossover exponent $\theta$ as deduced from (\ref{e5}). It is given by 
the slope of the curve.}  \label{simtheta}
\end{center}
\end{figure}
This exponent is determined by trial and error to get the maximum collapse of the data on a 
single curve. The figure shows indeed a nice data collapse but it hides a subtlety which 
we can uncover using the properties of the crossover function $g(x)$.
The function $g(x)$ should be expandable for small arguments as
\begin{equation} \label{e3}
g(x) = g_0 + g_1 x + \cdots
\end{equation} 
and for large arguments as
\begin{equation} \label{e4}
g(x) \simeq {1 \over x} \left( g_{-1} + {g_{-2} \over x} + \cdots \right).
\end{equation}
Then $\tau$ goes as $N^3$ for vanishing $c$ and as $N^2$ for $N \rightarrow \infty$ (at 
non-zero $c$). Inserting the asymptotic behavior (\ref{e4}) into (\ref{e1}) we obtain
\begin{equation} \label{e5}
\ln(\tau/N^2) = \ln g_{-1} - \theta \ln c + \cdots,
\end{equation} 
where the dots refer to corrections of order $1/N$. In Fig. \ref{simtheta} we have made a
plot of the limit of $\ln (\tau/N^2)$ vs $\ln c$. The values of the vertical
axis are extrapolated to $N \rightarrow \infty$, which 
corresponds to the first two terms of (\ref{e5}). We first check whether the basis of states 
is large enough, such that we have no systematic errors due to a too small basis. 
Then we inspect whether $\tau/N^2$ has a well defined limiting value for 
$N \rightarrow \infty$. (The values should approach the limiting point in a fairly linear
way.) If the curves in Fig. \ref{simtheta} had a straight slope, 
a well defined value of the crossover exponent $\theta$ follows. 
As one observes, there rather is a constant slope for small values of $c$ 
and another one for the larger values of $c$. Now crossover may only be expected in the
limit of $c \rightarrow 0$, which gives a slope in the neighborhood of $\theta \simeq 0.5$. 
We could therefore discard the behavior for larger $c$, as not being described by crossover, 
but this contrasts the findings for the cage model, where the crossover formula applies
for practically the whole range of $c$. We show the data also for larger values of $c$ because
we find it intriguing that this region is also representable by a crossover 
function, albeit with a different crossover exponent.      
\begin{figure}[ht]
\begin{center}
    \epsfxsize=12cm%\linewidth
    \epsffile{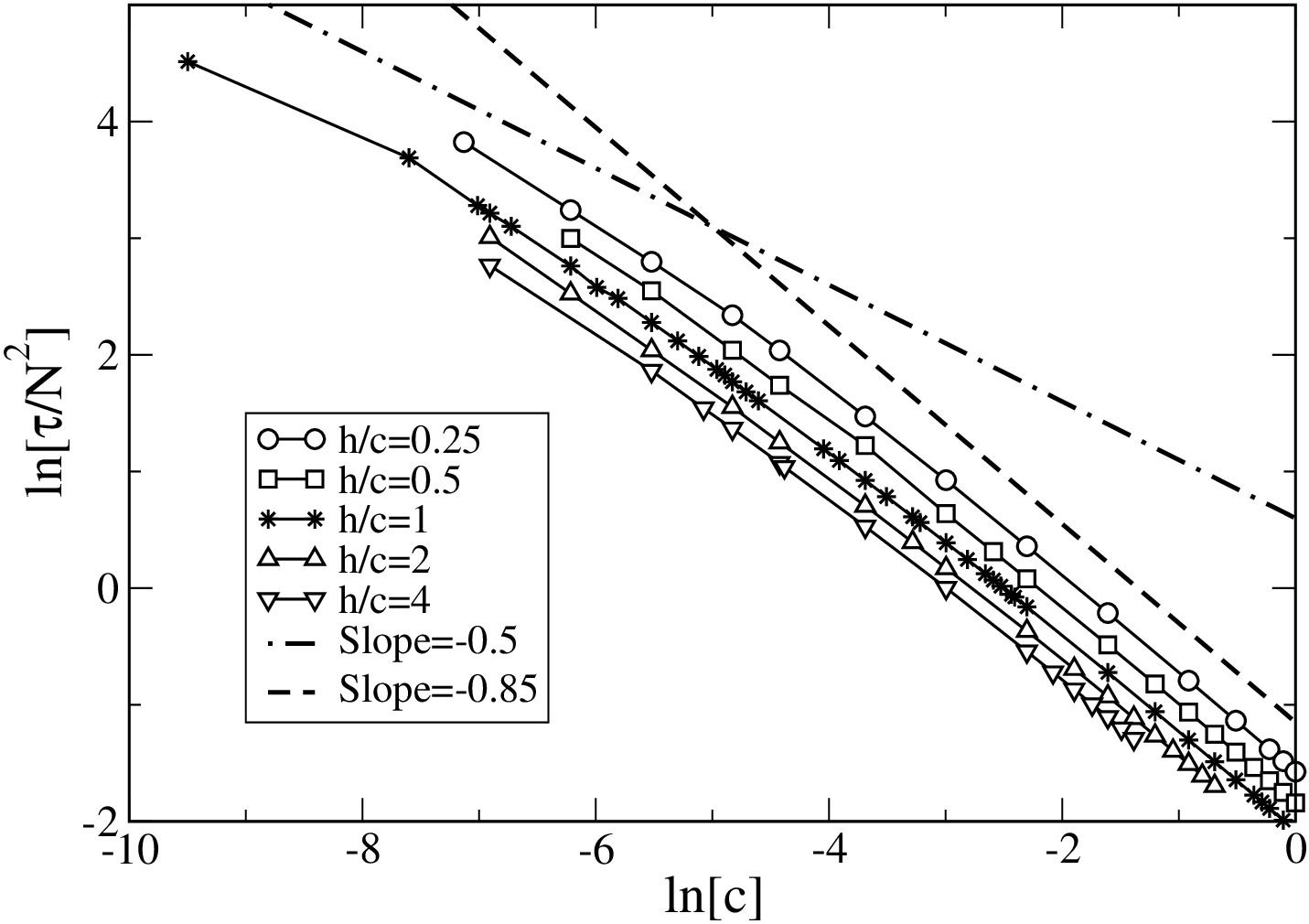}
   \vspace*{5mm}

    \caption{The crossover exponent $\theta$ for ratios of $h/c$.}  \label{mixtheta}
\end{center}
\end{figure}

\section{Crossover along lines $h/c=r$}

With one of the parameters $h$ or $c$ fixed it is the other parameter which controls
the crossover. The real challenge is to find a representation where both mechanisms
feature. We have not been able to find a simple expression which accurately accounts
for arbitrary combinations of $h$ and $c$. We gain some insight in the
combined action of $h$ and $c$ by approaching the limit $h=c=0$ along a radial line
$h/c=r$. For fixed $r$ we have again a single parameter which, in combination with $N$,
provides a crossover scaling variable, such that we can use 
the scenario of the previous section to analyze the data. 
In Fig. \ref{mixtheta} we give the crossover exponent $\theta$ as
a function of $\ln c, (\ln h)$ and for some values of the parameter ratio $h/c$.
For very small values of $\ln c, (\ln h)$ the slope of the line is compatible with the
``universal'' exponent $\theta =0.5$. However for larger values another exponent
seems to emerge of the order $\theta=0.85$. Note that the curves in  Fig. \ref{mixtheta}
run quite parallel, which means that $r$ only enters in the offset given by $g_{-1}$ in (\ref{e5}).
\begin{figure}[ht]
\begin{center}
    \epsfxsize=12cm%\linewidth
    \epsffile{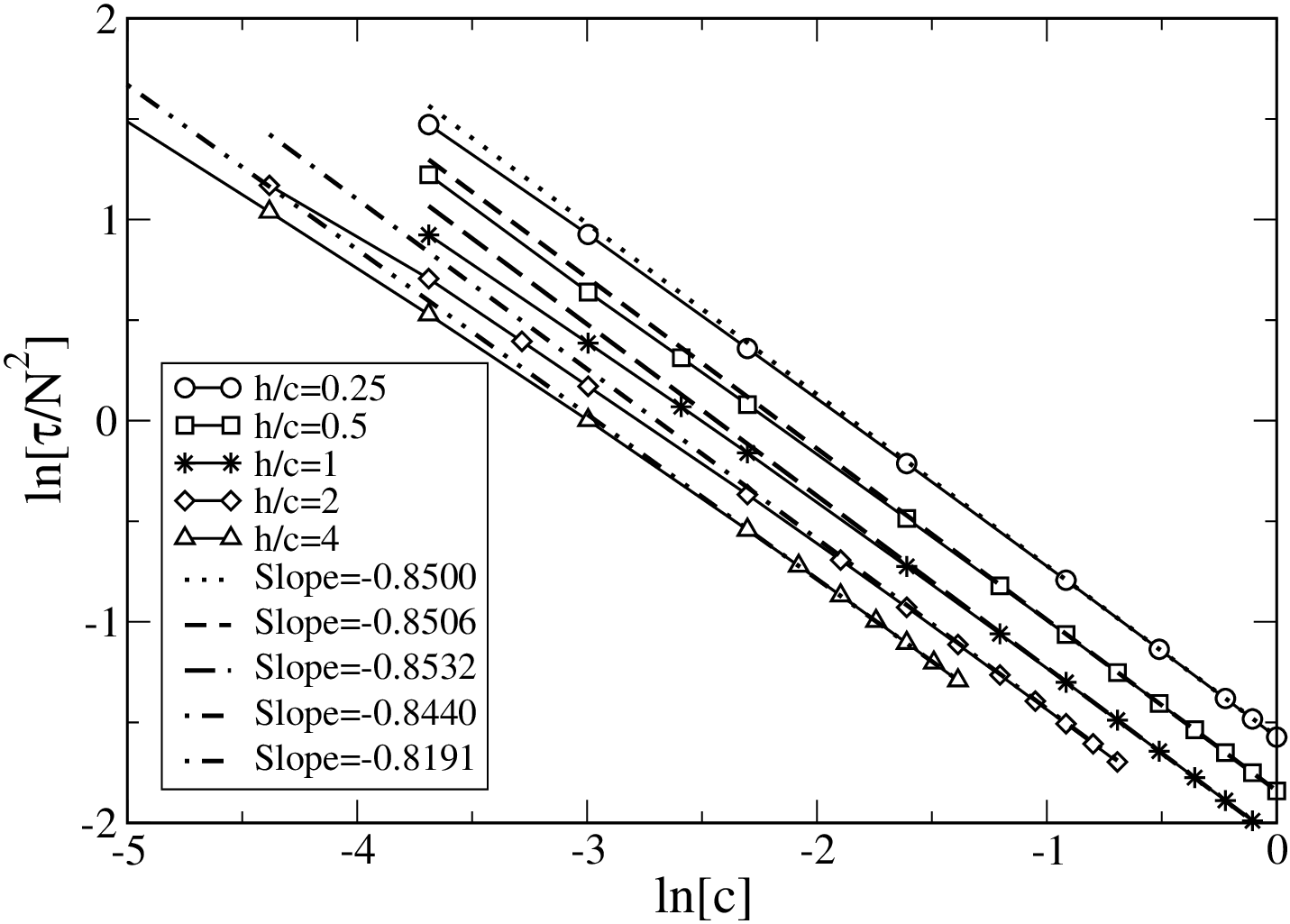}
   \vspace*{5mm}

    \caption{The crossover exponent $\theta$ for ratios of $h/c$.}  \label{anotheta}
\end{center}
\end{figure}
To show this point in more detail we have plotted in Fig. \ref{anotheta} the lines for 
larger values of $h$ (or $c$). The window where the large exponent $\theta = 0.85$ applies
is more than an order of magnitude for a fixed ratio $h/c$.  

\section{The Diffusion Coefficient D}

\begin{figure}[ht]
\begin{center}
    \epsfxsize=12cm%\linewidth
    \epsffile{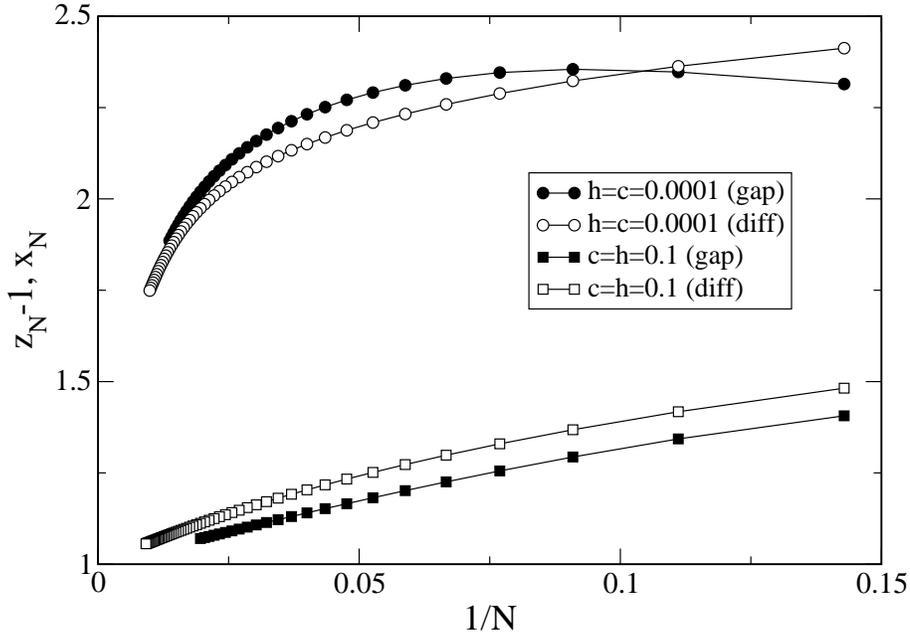}
   \vspace*{5mm}

    \caption{Comparison of the exponents $z_N-1$ and $x_N$.}  \label{compEXP}
\end{center}
\end{figure}
The diffusion coefficient has been determined by the linearization (\ref{a5}), which 
gives the linear response of the drift velocity with respect to the driving force. 
We do not repeat the analysis for the diffusion exponent $x$, since diffusion
and renewal time are closely related. If the center of the chain has drifted over a distance of
the order of the end-to-end distance $\sqrt{N}$, the chain has renewed itself. 
The mean square displacement due to diffusion during a renewal time equals $D \tau$. So 
one has the connection
\begin{equation} \label{f1}
D \tau \simeq N \quad \quad {\rm for} \quad \quad N \rightarrow \infty.
\end{equation} 
This relation implies for the exponents $z-x=1$. We have tested this relation and
in Fig. \ref{compEXP}  we show the values of $z_N - 1$ and $x_N$ for the same set of 
parameters, one for the small value $h=c=0.0001$, where the behavior is more reptative and
one for the larger value $h=c=0.1$, where the exponents tend to Rouse dynamics. See also
Fig. \ref{compd2d3} which shows that the crossover exponent $\theta$ for renewal and for 
diffusion are practically the same in the domain where it could be calculated with 
reasonable accuracy. 

\section{Two-dimensional results}

\begin{figure}[ht]
\begin{center}
    \epsfxsize=12cm%\linewidth
    \epsffile{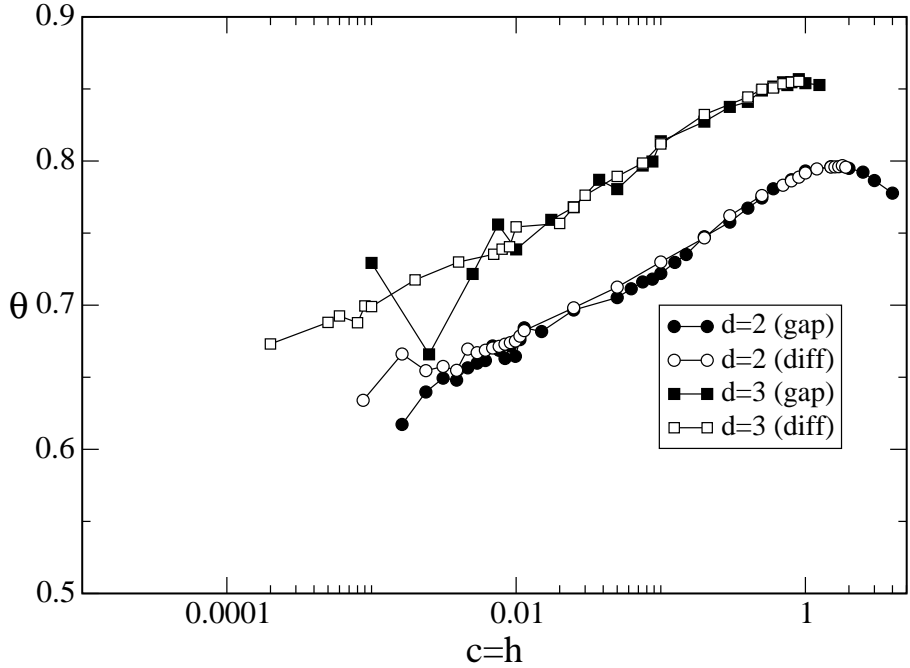}
   \vspace*{5mm}

    \caption{Comparison of the crossover exponent $\theta$ for $d=2$ and $d=3$, both for the gap and 
the diffusion coefficient.}  \label{compd2d3}
\end{center}
\end{figure}
We have shown the results for embedding dimension $d=3$. As mentioned 
the general expectation is that the embedding dimension has little influence on the
universal properties. We are now in a position to verify this statement, since we have
made extensive calculations both in $d=2$ and $d=3$.
Indeed we come to the conclusion that the results agree qualitatively. To show
an example we plot in Fig. \ref{compd2d3} the $d=3$ and $d=2$ curves for $\theta$ for $h=c$.
The trends are the same, but the value of $\theta$ in the ``large'' parameter regime is
definitely larger for $d=3$ than for $d=2$.

\section{Discussion}

As a follow up of the study of crossover in the cage model, the extended RD model gives
by and large the same picture: with growing length the chain crosses over from reptation
to Rouse dynamics. The story is here more complicated because the two extra types of hopping,
barrier crossing and hernia creation/annihilation, have to assist each other in order to
get Rouse dynamics for long chains. This makes a comprehensive representation of 
the data in one scaling expression complicated. We have investigated the crossover
behavior along lines in the $c,h$ plane. 

The underlying idea of crossover is that there are two competing time scales. One is the
diffusive time scale $N^2 /D_c$, which is the time needed for a perturbation to diffuse
along the chain inwards. We note that here not the overall diffusion $D$ but 
the curvilinear diffusion coefficient $D_c$ applies,
which decays as $N^{-1}$. This timescale leads to a renewal time $\sim N^3$. The other
time scale is the time needed to renew the chain by the combined action of hernia 
creation/annihilation and barrier crossing. If one of the parameters $c$ or $h$ is large 
enough, the other is the limiting factor. If $c$ is the smaller one, the timescale due to $c$
equals $N/c$ and if $h$ sets the rate, it is $N/h$. The fastest of the two time scales sets
the overall rate and therefore crossover occurs when they are equal i.e. when $c \sim N^{-2}$,
or $h \sim N^{-2}$, whichever is the smaller parameter. This leads to a crossover exponent 
$\theta =1/2$. We see this trend in the numerical data, but the fact that we have to go to 
really small values of $c$ (or $h$), and therefore to correspondingly large $N$, prevents
to see this ``universal'' crossover exponent in a clear way.  

On the other hand we observe for larger values of the parameters also crossover behavior,
with different crossover exponents $\theta$. This change in value could be a demonstration
of corrections to scaling, just as the exponents $z_N$ or $x_N$ are rather far from their
asymptotic values, also when crossover plays no role (as e.g. Fig. \ref{lackbar} shows).
Clearly the renewal of the chain by sideways motion is still slow enough,
even when $c$ and $h$ are of order unity, such that the competition with the diffusive
renewal determines the character of the dynamics.

As pointed out earlier for the cage model \cite{DvL1}, the crossover differs from 
the common scenario for polymer melts, where the crossover is in the opposite direction:
from Rouse dynamics to reptation \cite{melts}. In the melt reptation results for the longer chains
because the restriction in motion of the polymer, due to the presence of others,
becomes more severe the longer the polymer is. We have argued that such crossover in 
melts can be understood from sideways motion that have a rate depending on the length
of the chain. If the renewal time is taken as indicative for the lifetime of a barrier,
the sideways motion would have a rate $  \sim N^{-z_N}$ in the melt. The combined
scaling parameter $c N^2$ then would shrink as $N$ grows. Since we always find
$z_N > 2$ reptation prevails for long chains in the melt. One would have to do a self
consistent calculation, as carried out by Paessens and Sch\"utz, to make this argument
quantitative \cite{paessens}.

Paessens and Sch\"utz \cite{paessens} have also extended the RD model with rates that
depend on the length of the chain. Their aim is to see the influence of "constraint
release" on finite chains. The constraint release that they allow is in our language
a mix of hernia creation and annihilation and barrier crossing. But not all types of
barrier crossing that we allow, are permitted in their model. So, it is a bit difficult
to make a clear comparison between their findings and ours. The interesting point
of their calculation is the requirement of selfconsistency: the rates determine the renewal
time and the renewal time in turn influences the rates. To carry out this program
accurately within the DMRG method is one of the challenges for further research.

{\bf Acknowledgment} The authors have benefitted very much from many stimulating 
discussions with Gerard Barkema. A.D. thanks Wroclaw Centre
for Networking and Computing for access for their computing
facilities (grant No. 82).

\appendix 

\section{The Symmetries of the Master Operator}

It is easy to set up a DMRG without paying attention to the symmetries of the problem. 
Then in $d=3$ each link can be in $2d+1=7$ states, leading to $7^N$ configurations for
$N+1$ reptons. The possible symmetries in the problem will give an equal probability
to many configurations. If the symmetries are not explicitely acknowledged, the symmetries
get lost when the choice of basis states does not conserve the symmetry. This means that
states, which are equivalent by symmetry, have to be chosen simultaneously. 
Thus either one has to include a large number of states, which leads to impractical 
calculations, or one has to keep track of the symmetry in each step of the method,
which requires a substantial extra amount of careful programming. 
However, since we want to extract the utmost out of the data, we have 
no choice and must optimize the symmetry.

As we deal with the fieldless gap for the renewal time and with the fieldless equation
(\ref{a5}), we can employ, in principle, the full symmetry group of the cube, 
which has 48 elements. If we were to do an exact calculation, we could apply all the
relevant symmetry operations to the wavefunction and so reduce the number of 
components. But that is not the way DMRG works. The configuration space is splitted
into a tail and head part and the wavefunction is improved by an optimal choice of
basis states in one part using the density matrix induced by the other part. In order
to keep the symmetry in the wavefunction, the chosen states have to have the symmetry,
which implies that the density matrix must have the symmetry. That in turn implies 
that, in each stage of the calculation, the wavefunction of the whole chain must have
the desired symmetry. Thus we have to know how to combine symmetry of the parts in
order to get the symmetry of the whole. This is similar to combining angular momenta of
particles in atomic physics to get the angular momentum of the total wavefunction.
The ``good quantum numbers'' derive from a set of commuting symmetry operators.

We can find at most 3 commuting operators within the cubic group, with some freedom 
of choice. The simplest would be to look to the reflection symmetry of the coordinate axes. 
For each of the 3 operations the wavefunction can be even or odd, giving 8 sectors 
labeled by the parities. The parities qualify as good quantum numbers. 
The groundstate (stationary state) is located in the sector which is even under 
all three  reflections. Each of the other sectors contains an excited state and 
the smallest (in magnitude) is the gap. 
The parities of the parts can be easily be combined to parities for the total since they
simply multiply. We have implemented this scheme, but it does not lead to very 
accurate results, which we blame to the rather unbalanced occupation of the sectors,
when the most probable states of the density matrix are chosen. 

The most successful use of the symmetry comes from another choice of commuting symmetry
operations. We put the field in the direction of the body diagonal and consider 
rotations around this diagonal. We may rotate over the angles $\phi=0, 2 \pi /3$ or 
$4 \pi /3$, leaving the problem invariant. Under a rotation the wavefunction is multiplied 
with a phase factor $\exp (i \phi)$, which qualifies also as a good quantum number, 
leading to 3 sectors.
Rotations commute with simultaneous inversion of the coordinate axes, doubling each of the 3 
rotation sectors. As first 3 sectors we take those which are invariant under inversion
and the next 3 are odd under inversion. The groundstate is in the first and the gap is 
the fourth sector (invariant under rotation and odd under inversion). The advantage of
these good quantum numbers is that we can combine the quantum numbers of parts (by simple
multiplication) to the same set of quantum numbers for the combination. For instance a
part in sector 2 ($\phi = 2 \pi/3$) and one in sector 3 ($\phi = 4 \pi/3$) lead to a 
combination with $\phi = 0$, which is therefore in sector 1. This 
use of symmetry gives good results, but it is not yet optimal.

A refinement could be made by considering the interchange of the $x$ and the $y$ 
axis. This turns the first and fourth sector into itself and transforms sector
2 and 3 as well as 4 and 5 into each other. Although there is no good quantum number
associated with this operation, we could use this symmetry by splitting sector 1 and 4 into 
an even and odd part under the interchange. This leads to 8 ``channels'', which partly
coincide with the previous sectors. Combining a part in sector (channel) 2 with a part 
in 3 does give an overall state in sector 1, but one has to take even and odd combinations 
to get them in the even and odd channel corresponding to sector 1. This gives a substantial 
amount of extra programming in order to keep properly track of the channels. But the effort
is rewarded, as it improves the accuracy which is needed for the delicate cases of the 
parameter space. This choice of sectors (channels) is efficient because the states of
the density matrix, which are chosen as having the largest eigenvalues, are more or less
evenly distributed over the channels.


\begin{thebibliography}{99}
\bibitem{deGennes} P. G. de Gennes, J. Chem. Phys. {\bf 55}, 572 (1971);\\
P.~G.~de Gennes, {\it Scaling Concepts in Polymer Physics},
Cornell University Press, Ithaca, USA 1971.

\bibitem{Rubinstein} M.~Rubinstein, Phys.~Rev.~Lett. \textbf{59}, 1946 (1987).

\bibitem{Duke} T.~A.~J.~Duke, Phys.~Rev.~Lett. \textbf{62}, 2877 (1989).

\bibitem{Doi} M.~Doi and S.~F.~Edwards, {\it The Theory of Polymer Dynamics}
(Oxford University, New York, 1989).

\bibitem{Viovy} J.~L.~Viovy, Rev.~Mod.~Phys. \textbf{72}, 813 (2000).

\bibitem{White} S.~R.~White, Phys.~Rev.~Lett. {\bf 69}, 2863 (1992);\\
U.~Schollwoeck, Rev. Mod. Phys.  {\bf 77}, 259 (2005).

\bibitem{Widom} B.~Widom, J.-L.~Viovy, and A.~D.~Defontaines,
J.~Phys.~I \textbf{1}, 1759 (1991).

\bibitem{Carlon} E.~Carlon, A.~Drzewi\'{n}ski, and J.M.J.~van~Leeuwen, 
Phys.~Rev.~E {\bf 64}, 010801(R) (2001).

\bibitem{paessens} M. Paessens and G. M. Sch\"utz, Phys. Rev. E {\bf 66} 021806 (2002).

\bibitem{DvL1} A.~Drzewi\'{n}ski, and J.M.J.~van~Leeuwen, Phys.~Rev.~E {\bf 74},
061801 (2006).

\bibitem{melts} Kurt Kremer, Gary S. Grest, and I. Carmesin, Phys. Rev. Lett.
{\bf 61}, 566 (1988);\\
A.~Wischnewski, M.~Monkenbusch, L.~Willner,
D.~Richter, and G.~Kali, Phys.~Rev.~Lett. {\bf 90}, 058302 (2003).

\end{thebibliography}
\end{document}